\documentclass[aps,prl,a4paper,twocolumn,longbibliography,superscriptaddress]{revtex4-1}
\usepackage{amsmath}
\usepackage{amssymb}
\usepackage{mathtools}
\usepackage{graphics}
\usepackage{comment}
\usepackage{tikz}

\newcommand{\beq}{\begin{equation}}
\newcommand{\eeq}{\end{equation}}
\newcommand{\bei}{\begin{itemize}}
\newcommand{\eei}{\end{itemize}}
\newcommand{\ben}{\begin{enumerate}}
\newcommand{\een}{\end{enumerate}}

\newcommand{\be}{{\mathbf e}}

\usepackage{hyperref}
\hypersetup{
   pdfpagemode=None, 
   pdfstartpage=1,
   pdfmenubar=true,
   pdftoolbar=true,
   colorlinks = true,
   linkcolor=blue,
   citecolor=blue,
   urlcolor=blue,
   bookmarksopen=false
 }

\definecolor{darkblue}{rgb}{0.,0.24,0.51}
\definecolor{britishracinggreen}{rgb}{0.0, 0.26, 0.15}
\definecolor{darkgreen}{rgb}{0,0.60,.2}

\def\be{\begin{equation}}
\def\ee{\end{equation}}

\def\rf#1{(\ref{#1})}

\begin{document}
\title{Universal aspects of a strongly interacting impurity in a dilute Bose condensate}
\author{Pietro Massignan}
\affiliation{Departament de F\'isica, Universitat Polit\`ecnica de Catalunya, Campus Nord B4-B5, E-08034 Barcelona, Spain}
\author{Nikolay Yegovtsev}
\author{Victor Gurarie}
\affiliation{Department of Physics and Center for Theory of Quantum Matter, University of Colorado, Boulder CO 80309, USA}

	\date{\today}
	
\begin{abstract}
We study the properties of an impurity immersed in a weakly interacting Bose gas, i.e., of a Bose polaron. In the perturbatively-tractable of limit weak impurity-boson interactions many of its properties are known to depend only on the scattering length. 
Here we demonstrate that for strong (unitary) impurity-boson interactions all static quasiproperties of a Bose polaron in a dilute Bose gas, such as its energy, its residue, its Tan's contact and the number of bosons trapped nearby the impurity, depend on the impurity-boson potential via a single parameter. 
\end{abstract}
\maketitle

\begin{figure}[b]
\includegraphics[width=\columnwidth]{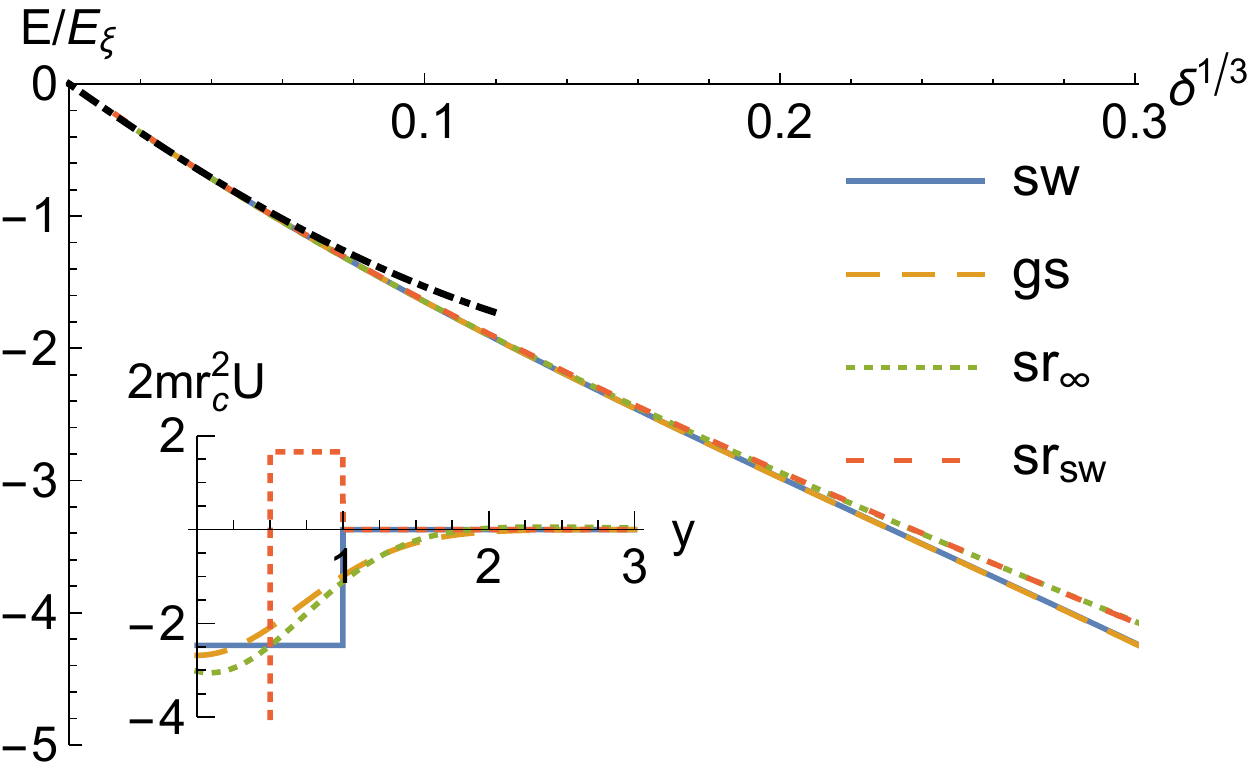}
\caption{\label{fig:energy}
\textbf{Polaron energy $E$ at unitarity} obtained numerically from various impurity-bath potentials tuned to their first unitary point: square well (sw), gaussian (gs), and shape-resonant (sr$_\infty$ and sr$_{\rm sw}$, which are respectively infinite- and finite-ranged; both have $r_e=0$)~\cite{potentialDefinitions}. 
Plot as a function of $\delta^{1/3}=(R/\xi)^{1/3}$ in units of $E_\xi = \xi n_0/(2m)$. 
The dot-dashed black line is our analytic result, Eq.~\eqref{eq:energyFinal}. The inset is a sketch of the potentials $U(y)$.
}
\end{figure}

The study of impurities in Bose and Fermi gases is an old and important subject \cite{Landau1933,Froehlich1954,Feynman1955,Gross1962,Padmore1971,Astrakharchik2004}. 
When a distinguishable atom is added to the gas, the impurity gets dressed by excitations in the bath and forms a quasiparticle often referred to as polaron.
Polarons in ultracold Fermi gases, both weakly and strongly interacting, have been studied for over a decade and a half \cite{Schirotzek2009,Chevy2010,Kohstall2012,Koschorreck2012,Massignan_Zaccanti_Bruun,Cetina2016,Scazza2017,Schmidt2018,Yan2018,Adlong2020}. 
The study of polarons in Bose gases picked up relatively recently, with the cases of neutral \cite{Cucchietti2006,Rath2013,Ardila2015,Christensen2015,Hu2016,Jorgensen2016,Grusdt2017,Yoshida2018,Yan2020,Drescher2020,Guenther2020}, Rydberg \cite{Balewski2013,Schlagmueller2016,Schmidt2016,Camargo2018} and charged impurities \cite{Cote2002,Massignan2005,Tomza2019,Astrakharchik2020,Dieterle2020} studied in the literature. 
Here we point out that an impurity added to a weakly interacting three-dimensional Bose gas exhibits universal features which depend very weakly on the details of the interaction potential (see Fig.~\ref{fig:energy}). Quite remarkably, we prove that the main features of Bose polaron can be calculated analytically even when the impurity interacts strongly, in the so-called unitary limit, with an otherwise weakly interacting Bose gas.

We consider a weakly interacting Bose gas made of atoms of mass $m$ with density $n_0$. 
Trading the density $n_0$ for the chemical potential $\mu$, we denote by $E(\mu)$ the energy cost of introducing a single impurity to the gas kept at this chemical potential. An impurity traps or repels extra bosons in its vicinity. It is possible to show that the number of trapped bosons is \cite{Massignan2005}
\be \label{eq:trapped} N = - \partial E /\partial \mu.
\ee
Our main goal is to calculate $E$, thus also determining the number of trapped bosons according to Eq.~\rf{eq:trapped}. In this work we limit ourselves to zero temperature. 

To describe a single heavy impurity, we can think of it simply as a radially-symmetric potential $U({r})$ it induces on the gas. 
The Hamiltonian ${\cal H}$ of the gas with an impurity is given by   (here and throughout we set $\hbar=1$)
\be \label{eq:ham} {\cal H} = \int d^3 x \left[ \frac{\nabla \bar \psi \nabla \psi}{2m} + \left[U(r) - \mu \right]\bar \psi \psi + \frac{\lambda}{2} \left( \bar \psi \psi \right)^2 \right],
\ee
where $m$ is the mass of the bosons. For mobile impurities with mass $M$, in the latter equation we could replace $m$ with  the reduced mass $m_r = m M/(m+M)$ \cite{Gross1962, Guenther2020}. However, to simplify equations, we will limit ourselves to the case $M \gg m$, so that $m_r=m$. 
The coupling constant $\lambda$ is related to the scattering length $a_B>0$ characterizing interactions among bosons by $\lambda =  4 \pi a_B/m$. 
The gas we consider here is weakly-interacting, which is well known to imply that
\be \label{eq:weakc} n_0 a_B^3 \ll 1. \ee 
The chemical potential $\mu$ can be used to define the healing length $\xi$ of the gas according to
\be \mu = 1/\left( 2 m \xi^2 \right).
\ee
In a weakly-interacting Bose gas $\mu = \lambda n_0$, and the condition for weak interactions can be rewritten as \be \label{eq:weakcc} \xi \gg n_0^{-1/3} \gg a_B, \ee where  $n_0^{-1/3}$ is the mean interparticle spacing in the gas. 

Suppose the impurity-bath potential $U(r)$ is characterized by a scattering length $a$. It has been known for quite some time that, for small enough $\left| a \right|$, the polaron energy and the number of trapped bosons are given by
\be \label{eq:weak} E = 2 \pi n_0 a/m, \qquad N = -a/(2 a_B).
\ee
For an attractive potential $E<0$ and $N>0$, since $a<0$.

The results in Eq.~\rf{eq:weak} are well known, yet it has not been known how small $\left| a \right|$ should be in order for it  to hold, nor has been known what it gets replaced by as the potential is made progressively more attractive towards the unitary point $1/a=0$. Here we show that the appropriate conditions, as well as all polaron quasiparticles at unitarity, can be stated in terms of the range of the potential $R$. While it might be intuitively clear what the range of the potential represents, we need to define $R$ precisely. In our context, this will be done using the following construction. 
Consider the zero energy Schr\"odinger equation 
\be \label{eq:schrr} - \frac{\Delta \psi_0 }{2 m} + U\psi_0 =0
\ee
with the potential $U$ tuned to unitarity (for example, by varying its amplitude). Its solution $\psi_0$ must go as $1/r$ at large distances, at least for potentials which vanish faster than $1/r^2$. 
Let us normalize the solution by demanding that $r \psi_0 \rightarrow 1$ as $r$ goes to infinity.
We now define $R$ as
\be \label{eq:defR} R \equiv \left[ \int \frac{d^3 x}{4\pi}  \left| \psi_0 \right|^4 \right]^{-1}.
\ee 
For potentials with a finite extent, $R$ can be shown to be close to their physical range. Often $R$ is also not far from the usual ``effective range" $r_e$ characterizing low-energy two-body scattering.
For example, for a unitary square well of width $r_c$ one has $r_e=r_c$ and $R/r_c=4/[2\pi{\rm Si}(\pi)-\pi{\rm Si}(2\pi)]\approx 0.557$, where ${\rm Si}(x)$ is the ``sine integral". 
We will return to this important point below.

It turns out, as we will show, that for the purpose of solving the polaron problem the potentials can be split into those which satisfy the following condition
\be \label{eq:rco} R \gg a_B \left( n_0 a_B^3 \right)^{1/4},\ee
and those which do not. 
In a weakly-interacting bath $a_B$ is of the order of the range of the interaction potential among bath particles, and the latter is expected to be comparable to the range of the interactions with the impurity, so that $R \sim a_B$. Taking into account the condition \rf{eq:weakc},  Eq.~\rf{eq:rco} is generally comfortably fulfilled in the atomic gases of interest here \cite{ultrashort}.  Note that this still implies $R \ll \xi$ thanks to Eq.~\rf{eq:weakcc}.

With this setup in place, we now present our results for interacting impurities satisfying Eq.~\rf{eq:rco}. The expression \rf{eq:weak} is only valid if
\be \label{eq:condition} \left| a \right|^3 \ll \xi^2 R.
\ee
As the potential $U$ grows more  attractive, $\left| a \right|$ grows. When the condition \rf{eq:condition} is violated, Eq.~\rf{eq:weak} breaks down. 
Exactly at unitarity where $a=\infty$, and in the limit  $R\ll \xi$, the energy of the polaron and the number of bath bosons in its dressing cloud can be found analytically and are given by
\be  \label{eq:EandN} 
E = - \frac{ 3 \pi n_0 \xi}{m} \left(  R/ \xi \right)^{1/3}, \quad N = 4 \pi n_0 \xi^3 \left( R / \xi \right)^{1/3}.
\ee
These constitute the most important results of this Letter. 
$N$ follows from $E$ in accordance with Eq.~\rf{eq:trapped}, where $\xi$ and $n_0$ have to be traded for $\mu$ before differentiating.
These results constitute the leading asymptotics of the solution when $R/\xi \ll 1$. Systematic higher order corrections to Eqs.~\rf{eq:EandN} are discussed below. 

To obtain these results we treat the Hamiltonian \rf{eq:ham} classically and solve the mean-field time-independent Gross-Pitaevskii (GP) equation instead of working with the fully quantum problem. 
One might worry that the GP equation is applicable only far away from the impurity where the condition \rf{eq:weakc} holds, but not nearby an attractively interacting impurity, where the local density of the gas $n_l$ is substantially higher than $n_0$. 
We will see later however that at unitarity $n_l = n_0 (\xi/R)^{4/3}$, thus the condition $n_l a_B^3 \ll 1$ is equivalent to Eq.~\rf{eq:rco} which, as we already discussed, we expect to hold true. 

The GP equation reads
\be \label{eq:GP} 
- \frac{\Delta \psi} {2m} + U \psi + \lambda \left| \psi \right|^2 \psi= \mu \psi.
\ee
Given the solution of this equation $\psi$, the energy of the polaron can be deduced by the substitution of it into Eq.~\rf{eq:ham} and subtracting the energy of the condensate without impurity, to give
\be \label{eq:energy1} 
E = -\frac{\lambda}{2} \int d^3 x  \left( \left| \psi \right|^4 - n_0^2 \right).
\ee
At the same time, the number of particles trapped in the polaron can be found by evaluating
\be N = \int d^3 x \left[ \left| \psi \right|^2 - n_0 \right].
\ee
We note that if the potential does not vary much on the scale of $\xi$, then the GP equation can be solved using local density approximation, as is often done in case when $U$ represents the smooth potential of a trap holding the condensate. However, we are interested in the opposite limit where the range of the potential is much smaller than $\xi$. Eq.~\rf{eq:GP} is nonlinear and at a first glance looks intractable. We now demonstrate that nevertheless its analytic solution is possible as long as $R \ll \xi$. 
 
We would first like to work with a potential which is strictly zero beyond some length $r_c$, $U(r)=0$ for $r>r_c$. Later we will show that it is also possible to consider potentials extending all the way to infinity. $R$ introduced above is of the order of $r_c$ but is not necessarily equal to it. We introduce $\phi = \psi/\sqrt{n_0}$.
Since we are looking for the lowest energy solution, this will be real valued and spherically symmetric. 

We analyze the Eq.~\rf{eq:GP} by introducing a small parameter $\epsilon = r_c/\xi$ and constructing its solution as an expansion in powers of this parameter.
As a first step, it is convenient to split the range of $r$ into $0 \le r \le r_c$ and $r_c \le r < \infty$. In the first interval we introduce $y = r/r_c$ and $\phi = \chi(y)/y$. $\chi(y)$ satisfies
\be \label{eq:GPrad1} 
-\frac{d^2 \chi}{dy^2}  + 2m r_c^2 U \chi = \epsilon^2 \left( \chi - \frac{\chi^3}{y^2} \right),
\ee
In the second interval we introduce $z=r/\xi$ and $\phi = 1+u(z)/z$, to find
\be \label{eq:GPrad2} -\frac{d^2 u}{dz^2} -2u = 3 \frac{u^2}{z} + \frac{u^3}{z^2},
\ee
where $u \rightarrow 0$ when $z \rightarrow \infty$. 
We need to solve Eqs.~\rf{eq:GPrad1} and \rf{eq:GPrad2}, matching the solutions at the boundary $r=r_c$. 

Now we will sketch the steps needed to follow through with this strategy, leaving details for Supplemental Material \cite{SuppMat}. Let us first examine the case of weakly attractive potential with a small scattering length $a<0$. We solve Eq.~\rf{eq:GPrad1} in the interval $0 \le y \le 1$ neglecting its right hand side as it contains a small parameter $\epsilon$. Then Eq.~\rf{eq:GPrad1} reduces to a Schr\"odinger equation in the potential $U$ at zero energy, whose solution $\chi_0$ must satisfy $\chi_0(0)=0$. We normalize the solution so that $\chi_0(1)=\alpha$, then $\chi'_0(1)$ satisfies Bethe-Peierls boundary conditions   $\chi'_0(1) = \alpha/(1-a/r_c)$. 

Now we solve Eq.~\rf{eq:GPrad2} neglecting its right hand side  to find $ u = A e^{-\sqrt{2} z}.$   This would be valid as long as $A$ is small. 
Matching amplitudes and derivatives of $\chi(y)$ and $u(z)$ at $z=\epsilon$, or correspondingly $y=1$, produces $ A = \epsilon(\alpha-1), \ -\sqrt{2} A = {\alpha}/\left( {1-a/r_c}  \right) - 1$. 
Taking into account that $\epsilon \ll 1$, this gives $A = -{a}/{\xi}$, and $\alpha = 1-{a}/{r_c}$. 
We can now plug our solution into Eq.~\rf{eq:energy1} and recover the answer \rf{eq:weak} under the condition \rf{eq:condition}. 

Let us now examine if the terms neglected to arrive at this solution are indeed small. We solve \rf{eq:GPrad1} by successive approximations, plugging $\chi_0$ into the right hand side of Eq.~\rf{eq:GPrad1} and producing a correction $\chi_1$. 
If $\left| a \right| < r_0$ then both $\chi_0(1)$ and $\chi'_0(1)$ are of the order of $1$ while $\chi_1$ will be of the order of $\epsilon^2 \ll 1$ and can be neglected. It gets more interesting if $\left| a \right| >r_c$. 
Then  $\chi_0(1) =\alpha \sim a/r_0$, while $\chi'_0(1) \sim 1$. At the same time, $\chi_1 \sim \epsilon^2 (a/r_c)^3$. The magnitude of this had better be smaller than $1$, so that the contribution of $\chi'_1(1)$ to the derivative of $\chi$ could be neglected. 
This condition gives $\epsilon^2 {\left| a\right|^3}/{r_c^3} \ll 1$. We thus recover the condition \rf{eq:condition} for the scattering length $a$ to be small enough so that the weak coupling solution is valid. Note that $A \ll 1$ and the right hand side of Eq.~\rf{eq:GPrad2} could indeed be neglected.

Suppose now that the potential $U$ is made more attractive so that its scattering length increases, violating the condition \rf{eq:condition} and eventually reaching infinity at the unitary point. We can follow the same strategy to obtain the solution in this case. The new element  is that Bethe-Peierls boundary conditions now imply $\chi'_0(1)=0$, so we need to solve Eq.~\rf{eq:GPrad1} perturbatively, using its right hand side as a perturbation, to find nonzero $\chi'_0(1)$. Same goes for Eq.~\rf{eq:GPrad2}. Leaving the details of the calculation for Supplemental Material \cite{SuppMat}, we find $\alpha = A/\epsilon$ with
\be A^3 = \epsilon \left(1+\int_0^1 dy \, v^4/y^2 \right)^{-1}.
\ee
Here $\psi_0 = v(y)/y$ is the solution of the Schr\"odinger equation at unitarity,  Eq.~\rf{eq:schrr}, normalized so that $v(1)=1$  ($y=1$ corresponds to $r=r_c$). 

Note that $\alpha \approx 1/\epsilon^{2/3}$ controls the amplitude of the solution to the Gross-Pitaevskii equation at $r<r_0$. It follows that the density of the gas at $r<r_0$ is roughly $n_l  \sim n_0 \alpha^2 =n_0/\epsilon^{4/3}$. This was used earlier to argue that $n_l a_B^3 \ll 1$ (or in other words $n(r)a_B^3\ll1$ for every $r$) implies Eq.~\rf{eq:rco}.

\begin{figure}
\centering
\includegraphics[width=\columnwidth]{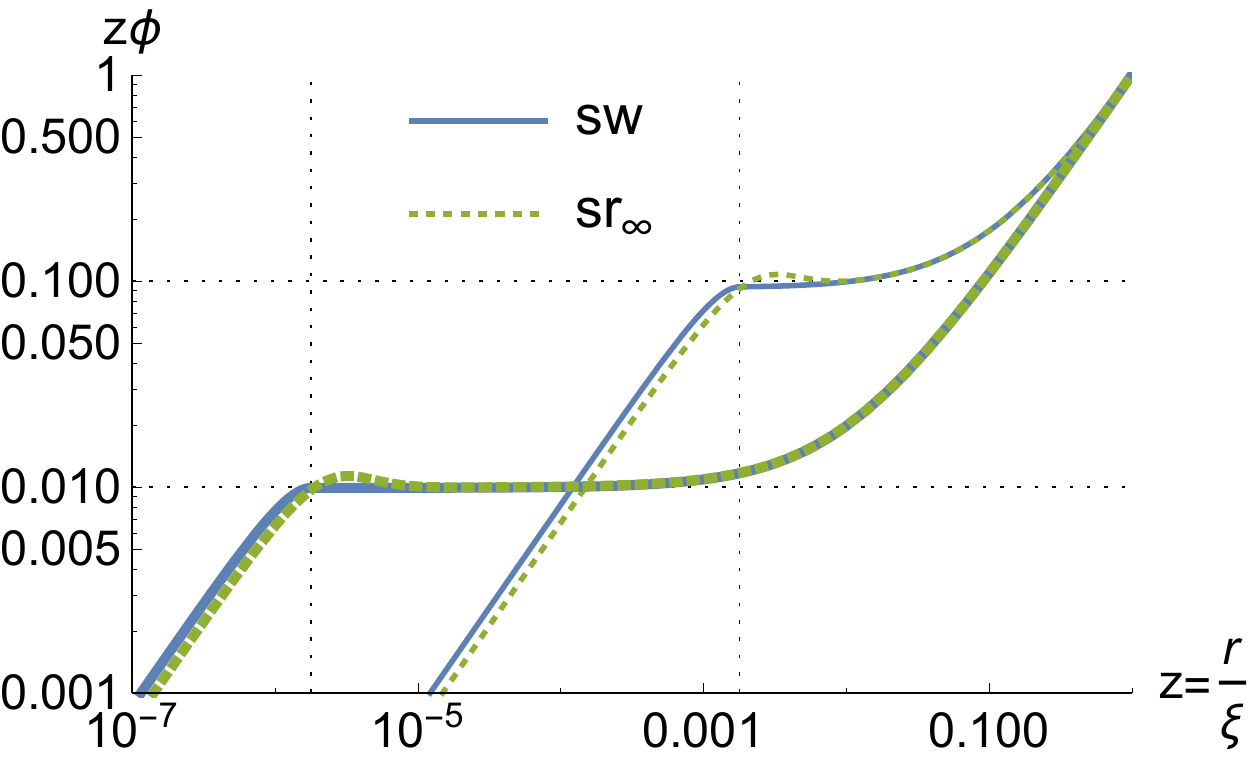}
\caption{\label{fig:wavefunctions}
{\bf Short-distance behavior.} Wavefunction $z \phi$ obtained from two unitary potentials (sw and sr$_\infty$) at $\delta=10^{-3}$ (thin) and $\delta=10^{-6}$ (thick). 
The horizontal and vertical dotted lines denote respectively the radii $r_c$ of the square well potentials, and the corresponding predictions $z\phi|_{r=r_c}\approx\delta^{1/3}$. 
}
\end{figure}

We can further observe that $v(y>1)=1$, so that $A$ can be conveniently rewritten as $A=\left(R/ \xi  \right)^{1/3}$, where $R$ was defined in Eq.~\rf{eq:defR}. Note that $R$  can be computed even when the potential extends to infinity, so at this point we can safely take the limit $r_c \rightarrow \infty$ if desired. The algebra leading to this result is also presented in the Supplemental Material \cite{SuppMat}.
Wavefunctions obtained numerically for two different $U(r)$ are shown in Fig.~\ref{fig:wavefunctions}. Even though the potentials have very different features (sw is finite-ranged with effective range $r_e=r_c$, while sr$_\infty$ is infinite-ranged with $r_e=0$), the wavefunctions obtained at equal values of $\delta=R/\xi$ are remarkably similar.

We can now plug the solution thus computed  into Eq.~\rf{eq:energy1} to find the expression for the energy, which is Eq.~\rf{eq:EandN}.
The procedure outlined here can actually be further used to construct a perturbative expansion in powers of $\epsilon$ or, more precisely, $\delta = R/\xi$, which gives
\begin{eqnarray} \label{eq:energyFinal}
E&  = &  - \frac{ \pi n_0 \xi}{m} \left( 3 \delta^{\frac 1 3} - 2 \sqrt{2} \delta^{\frac 2 3} + 4 \delta \ln \delta + \dots \right), \cr
N & = & 4 \pi n_0 \xi^3 \left(  \delta^{1/3} - \frac{5}{3 \sqrt{2}} \delta^{2/3} + 2 \delta \ln \delta  + \dots \right).
\end{eqnarray}
Terms beyond those shown here will require constructing further perturbative expansion of Eq.~\rf{eq:GPrad1}, and are expected to be less universal, depending on the features of the potential beyond those controlled by $R$. 
In Fig.~\ref{fig:energy} we show that the numerical solution of the GP equation using various (finite- and infinite-ranged) unitary impurity-bath potentials $U(r)$~\cite{potentialDefinitions} yields polaron energies which are remarkably independent of $U(r)$, and are in very good agreement with our analytical result Eq.~\eqref{eq:energyFinal}. 

The residue $Z$ quantifies the overlap between the solutions in presence and absence of the impurity. Within the GP treatment, this is given by $\ln Z= - \int d^3x\, |\psi(x)-\sqrt{n_0}|^2$ \cite{Guenther2020}. At unitarity, the above analysis shows that to leading order $\ln Z=-\sqrt{2} \pi n_0\xi^3 \delta^{2/3}$.

Another key quasiparticle property is the impurity-bath Tan's contact, which quantifies the change in the polaron energy in response to a small change of the inverse scattering length, $C = -8 \pi m \,{\partial E}/{\partial a^{-1}} $. An alternative definition of the contact is based on the impurity-bath density-density correlator evaluated at the core radius, $\tilde C =  16 \pi^2 r_c^2 \left| \psi(r_c) \right|^2$. Our formalism allows us to compute both quantities, and we have directly verified that at unitarity an identical answer is obtained from both definitions:
\be \label{eq:contact} C= \tilde C = 16 \pi^2 n_0 \xi^2 \delta^{2/3}  \left( 1 - 2\sqrt{2} \delta^{1/3}/3 + \dots \right).
\ee
in the leading approximation in $\delta$. However, we have not established whether $C$ remains to be equal to $\tilde C$ in higher order terms in $\delta$, and neither are we  aware of a general argument establishing their equality. We also note that the definition $\tilde C$ given above stops working when $r_c$ is infinity, as is the case with infinite range potentials.

\begin{figure}
\centering
\includegraphics[width=\columnwidth]{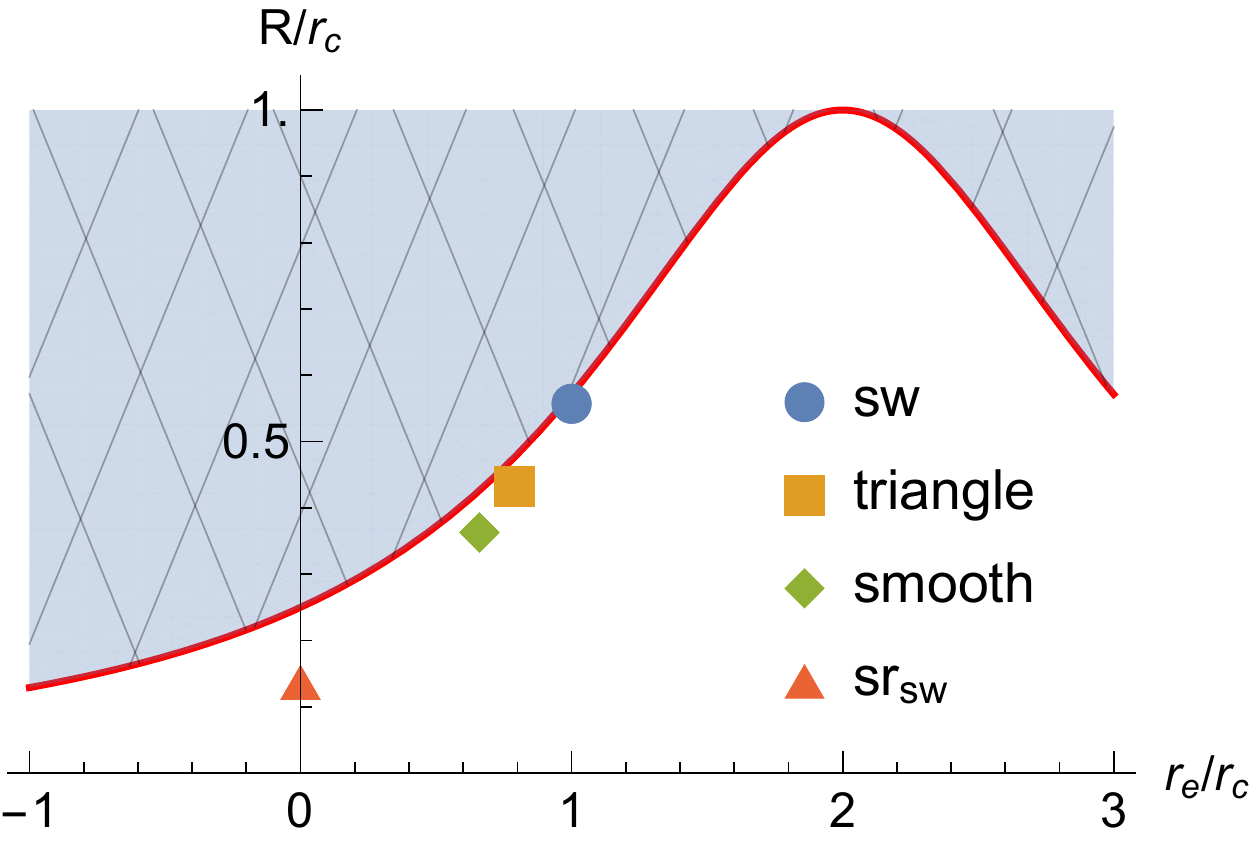}
\caption{\label{fig:bound}
{\bf Upper bound to $R$.} The range $R$ of potentials with a finite range (i.e., such that $U(r>r_c)=0$) is bound from above by a simple function of the effective range $r_e$ (red line), see Eq.~\eqref{bound}. The symbols show the range $R/r_c$ obtained from various finite-ranged potentials~\cite{potentialDefinitions}.
}
\end{figure}

To get a grasp on the physical meaning of $R$ for finite-ranged potentials, consider the inequality $\int_0^{r_c} r^2 dr \left(\gamma + v^2/r^2 \right)^2 \ge 0$, which obviously holds for every $\gamma$. Minimizing with respect to $\gamma$, and using that the effective range at unitarity is given by 
$r_e= 2 \int_0^{r_c} dr \left( 1-v^2 \right)$  \cite{LandauLifshitzBookQM}, 
we find
\be 
\label{bound}
\frac{r_c}{R} \geq  \frac{3 r_e^2}{4 r_c^2} - \frac{3 r_e}{r_c} + 4.
\ee
Quite remarkably, this bound is often approximately saturated by many interesting potentials, as we show in Fig.~\ref{fig:bound}. This gives a way to estimate $R$ starting from the knowledge of $r_e$, which is experimentally of easy access.

As the potential $U$ increases in strengths beyond the unitary limit, which implies that it now has a bound state with binding energy $-\nu = -1/(2 m a^2)$, $a$ becomes positive. If $a$ becomes sufficiently small so that
the relationship \rf{eq:condition} holds again, simple arguments give the energy and the number of trapped particles of the polaron as
\be \label{eq:bound}  E  \sim -m R^3 \nu^2/a_B, \qquad N \sim m R^3 \nu/a_B,
\ee 
where the precise coefficients now depend on the details of the potential $U$ \cite{subtlety}. 

Indeed, suppose $N$ bosons get trapped in this bound state, then the polaron energy is 
$ E = - N \nu + g N^2/2$, where the self-repulsion constant $g$ can be estimated as $g \sim \lambda /R^3$. Minimizing $E$ with respect to $N$ we find Eqs.~\rf{eq:bound}. 
This solution can also be obtained from the GP equation if one notes that it corresponds to the density of bosons being $n_l \sim N/R^3 \sim \nu/\lambda$, and that results in the nonlinear term in the GP equation $\lambda \left| \psi \right|^2 \psi \sim \nu \psi$, thus turning the GP equation into the Schr\"odinger equation at energy close to $-\nu$ (neglecting the small term proportional to $\mu$). Such solution of the GP equation, which would fix the coefficients in Eq.~\rf{eq:bound}, can only be found numerically and the answer will be highly dependent on the details of the potential $U$. It is easy to see that $n_l a_B^3 \sim  (a_B/a)^2 \ll 1$ justifying the use of the GP equation. 

In conclusion, we presented here the complete analytic solution of a challenging many-body problem, the one of describing an impurity in strong interaction with a very compressible Bose bath. Our formalism shall hold under typical experimental conditions found in Bose polaron experiments, and it allows to compute many relevant quasiparticle properties, like the energy, the number of trapped bosons, the residue and the contact. In agreement with earlier studies, we showed that a strong attractive interaction generates a macroscopic coherent dressing of the impurity, which gives rise to a bosonic version of the orthogonality catastrophe in the limit of an infinitely compressible bath.

\vspace{5mm}
\begin{acknowledgments}
We acknowledge inspiring and insightful discussions with G.~E.~Astrakharchik, J. Levinsen, M. Parish, and specially with N.~E.~Guenther, who provided critical insight in the initial stages of this work. 
P.M. acknowledges support by the Spanish MINECO (FIS2017-84114-C2-1-P), and EU FEDER Quantumcat. This work was supported by the Simons Collaboration on Ultra-Quantum Matter, which is a grant from the Simons Foundation (651440, VG, NY).
\end{acknowledgments}

\bibliography{UnitaryPolaron}

\begin{thebibliography}{41}
\bibitem [{pot()}]{potentialDefinitions}%
  \BibitemOpen
  \href@noop {} {}\bibinfo {note} {Throughout this paper we consider two-body
  potentials tuned at their first unitary point. Here we provide their precise
  functional forms $2mr_c^2U(y)$.\\ Some of those are finite-ranged; for
  example, square well: $-(\pi/2)^2\Theta(1-y)$; shape-resonant$_{\rm sw}$:
  $-12.6\,(y<0.5), 1.66\, (0.5<y<1), 0\, (y>1)$; triangle:
  $-7.84(1-y)\Theta(1-y)$; smooth: $4.51\{1-\tanh[\tan(\pi(2y
  -1)/2)]\}\Theta(1-y)$.\\ The other ones are instead infinite-ranged;
  gaussian: $-2.68e^{-y^2}$; shape-resonant$_\infty$:
  $-3.73e^{-y^2}+0.71e^{-y}$.\\ The two potentials termed ``shape-resonant"
  have a vanishing effective range $(r_e=0)$.}\BibitemShut {Stop}%
\bibitem [{ult()}]{ultrashort}%
  \BibitemOpen
  \href@noop {} {}\bibinfo {note} {Looking at potentials with a very small $R$
  violating the condition \rf{eq:rco} may still be interesting from a purely
  theoretical standpoint, but is out of the scope of the present work. We plan
  to do it elsewhere.}\BibitemShut {Stop}%
\bibitem [{sub()}]{subtlety}%
  \BibitemOpen
  \href@noop {} {}\bibinfo {note} {A subtlety in trying to use
  Eq.~\rf{eq:trapped} in Eq.~\rf{eq:bound} is that an additional term in $E$
  suppressed by a power of a factor of $\delta$ compared to what is presented
  in Eq.~\rf{eq:bound} is needed to recover the expression for
  $N$.}\BibitemShut {Stop}%
\end{thebibliography}

\onecolumngrid
\begin{center}
\newpage
\textbf{
Supplemental Material:\\[4mm]
\large Universal aspects of a strongly interacting impurity in a dilute Bose condensate}\\
\vspace{4mm}
Pietro Massignan,$^{1}$ N. Yegovtsev,$^2$ Victor Gurarie$^2$ \\
\vspace{2mm}
{\em \small
$^1$Departament de F\'isica, Universitat Polit\`ecnica de Catalunya, Campus Nord B4-B5, E-08034 Barcelona, Spain\\
$^2$Department of Physics and Center for Theory of Quantum Matter, University of Colorado, Boulder CO 80309, USA
}
\end{center} 

\setcounter{equation}{0}
\setcounter{figure}{0}
\setcounter{table}{0}
\setcounter{section}{0}
\setcounter{page}{1}
\makeatletter
\renewcommand{\theequation}{S.\arabic{equation}}
\renewcommand{\thefigure}{S\arabic{figure}}
\renewcommand{\thetable}{S\arabic{table}}
\renewcommand{\thesection}{S.\arabic{section}}

\section{Solution of the Gross-Pitaevskii equation at unitarity\label{app:EnergyFunctional}}
We present here the details of the procedure leading to the analytical solution of the Gross-Pitaevskii equation
\be \label{eq:GPs} 
- \frac{\Delta \psi} {2m} + U \psi = \mu \psi - \lambda \left| \psi \right|^2 \psi.
\ee
Here $U(r)$ is an attractive potential tuned to unitarity. We will first assume that it has a finite range $r_c$ and vanishes if $r>r_c$, and relax this condition at the end of the calculation. The chemical potential is $\mu = 1/(2 m \xi^2)$ where $\xi$ is the healing length. We also have $\mu = n_0 \lambda$, where $\lambda =  4 \pi a_B/m$ and $n_0$ is the density of the condensate at $r \gg \xi$ where the condensate relaxes to its uniform state. As explained in the paper, we consider a very dilute bath, so that  $\xi \gg r_c$. 

We analyze Eq~\rf{eq:GPs} by introducing a small parameter $\epsilon = r_c/\xi \ll 1$ and constructing its solution as an expansion in powers of this parameter.
First it is convenient to introduce the condensate function normalized to $n_0$, 
\be \phi = \frac{\psi}{\sqrt{n_0}}.
\ee
We are looking for the lowest energy solution, so we expect $\phi$ to be real and spherically symmetric. 

Next, it is convenient to split the range of $r$ into $0 \le r \le r_c$ and $r_c \le r < \infty$. In the first interval we introduce $y = r/r_c$ and $\phi = \chi(y)/y$. $\chi(y)$ satisfies
\be \label{eq:GPrad1s} 
-\frac{d^2 \chi}{dy^2}  + 2m r_c^2 U \chi = \epsilon^2 \left( \chi - \frac{\chi^3}{y^2} \right),
\ee
In the second interval we introduce $z=r/\xi$ and $\phi = 1+u(z)/z$, to find
\be \label{eq:GPrad2s} -\frac{d^2 u}{dz^2} -2u = 3 \frac{u^2}{z} + \frac{u^3}{z^2},
\ee
where $u \rightarrow 0$ when $z \rightarrow \infty$. 
We need to solve Eqs.~\rf{eq:GPrad1s} and \rf{eq:GPrad2s}, matching the behavior of their solutions at the boundary $r=r_0$. 

First let us discuss Eq.~\rf{eq:GPrad1s}. As a zeroth approximation, we solve it without its right hand side, and then construct corrections to this by means of successive approximations in powers of $\epsilon$,
\be \chi = \chi_0 + \chi_1 + \dots.
\ee
At zero-th order, we need to solve
\be -\frac{d^2 \chi_0}{dy^2}  + 2m r_c^2 U \chi_0 =0.
\ee
This is the Schr\"odinger equation in the potential $U(r)$ tuned to unitarity. Its solution must satisfy Bethe-Peierls boundary conditions at $r=r_c$ or $y=1$, which for a potential at unitarity give  $\chi_0'(1)=0$. The amplitude of the solution is arbitrary at this point and will be determined by matching with the solution of Eq.~\rf{eq:GPrad2s}. This will show that the amplitude goes as $1/\epsilon^{2/3}$. In anticipation of this, let us introduce the following notation 
\be \chi_0(y) = \frac{\beta} {\epsilon^{2/3}} v(y). \end{equation} Here $v$ is the solution of the Schr\"odinger equation
\be -\frac{d^2 v}{dy^2}  + 2m r_c^2 U v =0
\ee
normalized so that $v(1)=1$. We have $v'(1)=0$ since $U$ is tuned to unitarity. $\beta$ is a yet unknown $\epsilon$-independent normalization coefficient (related to $\alpha$ introduced in the main text by $\beta= \alpha \epsilon^{2/3}$). 

We will need a correction to this which satisfies
\be 
\label{eq:p1} -\frac{d^2 \chi_1}{dy^2}  + 2m r_c^2 U \chi_1 = -\epsilon^2  \frac{\chi_0^3}{y^2}.
\ee
The term $\epsilon^2 \chi_0$ from the right hand side of Eq.~\rf{eq:GPrad2s} goes as $\epsilon^{4/3}$ and can be neglected. At the same time, we see that $\chi_1$ is of the order of $\epsilon^0$. Solving  Eq.~\rf{eq:p1} gives
\be \chi_1 = \beta^3 v(y) \int_0^y \frac{ds}{v^2(s)} \int_0^{s} \frac{dt \, v^4(t)}{{t}^2}.
\ee
Putting it together produces
\be \label{eq:seconds} \chi = \frac{\beta}{\epsilon^{2/3}}  v(y) + \beta^3 v(y) \int_0^y \frac{ds}{v^2(s)} \int_0^{s} \frac{dt \, v^4(t)}{{t}^2} + \dots.
\ee
The next term $\chi_2$ which can be obtained by continuing successive approximations goes as $\epsilon^{2/3}$. We will not need it here, but note that it will have an even more complicated dependence on $v$ and by extension on features of the potential $U(r)$ than the already obtained term $\chi_1$.

From this solution we find that
\be \label{eq:bb1} \chi(1) = \frac{\beta}{\epsilon^{2/3}} + \mathcal{O}(1),  \quad \chi'(1) = \beta^3 c + \mathcal{O}(\epsilon^{2/3}), \quad c= \int_0^1 \frac{dy \,
v^4(y)}{y^2}.
\ee

Now we turn our attention to Eq.~\rf{eq:GPrad2s}. Its solution $u(z)$ needs to be matched with the boundary conditions \rf{eq:bb1}. Easy to verify that these boundary conditions imply
\be \label{eq:bb2} u(\epsilon) = \beta \epsilon^{1/3} + \mathcal{O}(\epsilon), \quad u'(\epsilon) = - 1 + \beta^3 c + \mathcal{O}(\epsilon^{2/3}).
\ee

Eq.~\rf{eq:GPrad2s} differs from Eq.~\rf{eq:GPrad1s} in that its nonlinear terms do not have an explicit factor of $\epsilon$ in front of them. We will nonetheless solve 
Eq.~\rf{eq:GPrad2s} by means of successive approximations. Without its right hand side, the solution to Eq.~\rf{eq:GPrad2s} reads
\be \label{eq:uzero} u_0 (z) = A \, e^{- \sqrt{2} z}.
\ee
We use this to rewrite Eq.~\rf{eq:GPrad2s} as an integral equation via a standard procedure. This involves solving the auxiliary equation
\be -\frac{d^2 u}{dz^2} -2u = g(z)
\ee
with arbitrary given $g(z)$, then substituting the actual right hand side of Eq.~\rf{eq:GPrad2s}. We find
\be  \label{eq:it}  u(z) = A \, e^{-\sqrt{2} z} +  \frac{e^{-\sqrt{2} z}}{2 \sqrt{2}} \int_z^\infty \frac{ds \, e^{\sqrt{2} s}}{s} \left(3 u^2 (1+u') + \sqrt{2} u^3 \right) -  \frac{e^{\sqrt{2} z}}{2 \sqrt{2}} \int_z^\infty \frac{ds \, e^{-\sqrt{2} s}}{s} \left(3 u^2 (1+u') - \sqrt{2} u^3 \right). 
\ee
We now use this equation to calculate $u(\epsilon)$ and $u'(\epsilon)$ in perturbative expansion in powers of $\epsilon$. Anticipating that the leading behavior is $A \sim \epsilon^{1/3}$, as should be clear from comparing Eq.~\rf{eq:uzero} and Eq.~\rf{eq:bb2}, we iterate Eq.~\rf{eq:it} by plugging $u_0(z)$ into the right hand side of Eq.~\rf{eq:it}. The resulting integrals can be computed in terms of Gamma functions and expanded in powers of $\epsilon$.  This allows us to evaluate $u(\epsilon)$ to be
\be \label{eq:va} u(\epsilon) = A + \frac{3 \ln 3 }{2 \sqrt{2}} A^2 - A^3 \ln \epsilon + \mathcal{O}(\epsilon).
\ee
We also evaluate $u'(\epsilon)$. Differentiating Eq.~\rf{eq:it} gives
\be \label{eq:derr} u'(z) = -\sqrt{2} A e^{-\sqrt{2} z} - \frac{u^3}{z}- \frac{e^{-\sqrt{2} z}}{2} \int_z^\infty \frac{ds \, e^{\sqrt{2} s}}{s} \left(3 u^2 (1+u') + \sqrt{2} u^3 \right) -  \frac{e^{\sqrt{2} z}}{2 } \int_z^\infty \frac{ds \, e^{-\sqrt{2} s}}{s} \left(3 u^2 (1+u') - \sqrt{2} u^3 \right).
\ee
We can again substitute $u_0(z)$ into the integrals on the right hand side of Eq.~\rf{eq:derr}, to find
\be \label{eq:derr2} u'(\epsilon) = -\sqrt{2} A - \beta^3 + 3 A^2 \ln \epsilon + \mathcal{O}(\epsilon^{2/3}).
\ee 
Here we took advantage of the boundary conditions \rf{eq:bb2} which tell us that $u(\epsilon) = \beta \epsilon^{1/3}$ within the accuracy that we work with. 

Combining Eqs.~\rf{eq:va} and \rf{eq:derr2} with Eq.~\rf{eq:bb2} gives 
\be \label{eq:syss}  \begin{matrix} A + \frac{3 \ln 3 }{2 \sqrt{2}} A^2 - A^3 \ln \epsilon & = & \alpha \epsilon^{1/3}, \cr
-\sqrt{2} A - \beta^3 + 3 A^2 \ln \epsilon & = &  - 1 + \alpha^3 c. \end{matrix}
\ee
We now need to solve these equations for $A$ and $\beta$ perturbatively, in powers of $\epsilon$. Introduce 
\be \delta = \frac{\epsilon}{1+c}.
\ee The solution to Eq.~\rf{eq:syss} reads
\begin{eqnarray} \beta \epsilon^{1/3}  &=& \delta^{1/3} - \frac{\sqrt{2}}{3} \delta^{2/3} + \delta \ln \delta + \mathcal{O}(\delta), \cr
A & = & \delta^{1/3} - \left(  \frac{3 \ln 3}{2 \sqrt{2}} + \frac{\sqrt{2}}{3} \right) \delta^{2/3} + 2\delta \ln \delta + \mathcal{O} ( \delta). 
\end{eqnarray}
We can now use the parameters we obtained in this way to calculate the energy and the particle number of the polaron. It turns out to be  technically easier to calculate the particle number first and then use Eq.~\rf{eq:trapped} to find the energy, which is the strategy we will follow here.

The excess number of bath particles trapped around the impurity is given by
\be N = \int d^3 x  \left[ \left| \psi \right|^2 - n_0 \right] = 4 \pi n_0 \xi^3 \int_0^\infty z^2 dz \left[ \phi^2-1 \right].
\ee
It is natural to split the integral over $z$ into two intervals, from $0$ to $\epsilon$ and from $\epsilon$ to infinity. Now the contribution of the first  interval can be safely neglected. Indeed, it gives
\be \int_0^{\epsilon} z^2 dz \left[ \phi^2-1 \right] = \epsilon^3 \int_0^1 dy \left( \frac{\chi^2}{y^2}-1 \right) \sim \epsilon^{5/3}.
\ee
Here we used that $\chi(y) \sim 1/\epsilon^{2/3}$. This contribution is very small and exceeds the accuracy in $\epsilon$ with which we were doing our calculations. This also indicates that the bulk of the particles bound by the impurity are located farther than distance $r_c$ away from the impurity. 

The contribution of the second interval gives
\be \label{eq:pp3} \int_\epsilon^{\infty} z^2 dz \left( \left( 1+ \frac{u}{z} \right)^2 - 1\right).
\ee
To evaluate this integral we again iterate  Eq.~\rf{eq:it} once, to find $u$ up to the terms of the order of $A^2$, and substitute that into Eq.~\rf{eq:pp3}. The result is
\be N = 4 \pi n_0 \xi^3 \left(  \delta^{1/3} - \frac{5}{3 \sqrt{2}} \delta^{2/3} + 2 \delta \ln \delta  + \dots \right).
\ee
Thus we evaluated the number of particles trapped in the polaron up to terms of the order of $\delta \ln \delta$. To go beyond this order, starting from terms of the order of $\delta$ and beyond represented by the dots above, we would need to go beyond the terms presented in Eq.~\rf{eq:seconds}. We expect that this will produce terms which depend on the features of the potential other than the coefficient $c$ (leading to $R$ from Eq.~\rf{eq:defR}, as we explain below). 

To construct the energy of the polaron, it is easiest at this stage to take advantage of Eq.~\rf{eq:trapped}. The subtlety in evaluating the derivative there is that the particle number $n_0$ as well as $\xi$ have to be traded for $\mu$ before differentiating. Doing the algebra we arrive at 
\be\label{eq:energyFinals}
E = - \frac{ \pi n_0 \xi}{m} \left( 3 \delta^{\frac 1 3} - 2 \sqrt{2} \delta^{\frac 2 3} + 4 \delta \ln \delta + \dots \right),
\ee
which is the same as Eq.~\rf{eq:energyFinal}.

Finally, let us examine $\delta= \epsilon/(1+c)$ in a little more detail. From the definition of $c$ given in Eq.~\rf{eq:bb1} we can write 
\be 1+ c = 1+ r_c \int_0^{r_0} \frac{dr \, v^4}{r^2} =r_c  \left( \int_{r_c}^\infty \frac{dr} {r^2} + \int_{0}^{r_c} \frac{dr \, v^4}{r^2} \right).
\ee
 $v$ is the solution of the Schr\"odinger equation with the potential tuned to unitarity, so that $v'(r_c)=0$. Since it is normalized such that $v(r_c)=1$, it will naturally satisfy $v(r)=1$ for all $r\ge r_c$. 
Therefore we can rewrite this as
\be 1+c = r_c \int_0^\infty \frac{dr \, v^4}{r^2}.
\ee
Now
\be \delta= \frac{\epsilon}{1+c} = \frac{1}{\xi  \int_0^\infty \frac{dr \, v^4}{r^2} }.
\ee
It is now convenient to define
\be R^{-1} =  \int_0^\infty \frac{dr \, v^4}{r^2}  = \int \frac{d^3 x} { 4 \pi}  \,  \left| \psi_0 \right|^4, 
\ee
where $\psi_0 = v/r$ is the solution of the Eq.~\rf{eq:schrr}. Thus we recover the definition
\be \delta = \frac{R}{\xi}
\ee
given in the main text, as well as the definition of $R$ given in Eq.~\rf{eq:defR}. At this stage $r_c$ drops from the equations and no longer needs to be finite. It can be taken to infinity if desired. 

\end{document}